\documentclass[aps,prb,amsmath,amssymb,footinbib,showpacs,twocolumn,superscriptaddress]{revtex4-2}
\usepackage{amsmath}
\usepackage{amssymb}
\usepackage{amsthm}
\usepackage{setspace}
\usepackage{graphicx}
\usepackage{braket}
\usepackage{mathrsfs}
\usepackage{float}
\usepackage[colorlinks = true,linkcolor = red,urlcolor  = blue,citecolor = blue,anchorcolor = blue]{hyperref}
\usepackage[utf8]{inputenc}
\usepackage[english]{babel}
\usepackage{bm}
\usepackage{comment}

\begin{document}

\title{Partially-separated Majorana modes in a disordered medium}

\author{Chuanchang Zeng}
\affiliation{Centre for Quantum Physics, Key Laboratory of Advanced Optoelectronic Quantum Architecture and Measurement(MOE),
School of Physics, Beijing Institute of Technology, Beijing, 100081, China}
\affiliation{Beijing Key Lab of Nanophotonics $\&$ Ultrafine Optoelectronic Systems,
School of Physics, Beijing Institute of Technology, Beijing, 100081, China}
\author{Gargee Sharma}
\affiliation{School of Basic Sciences, Indian Institute of Technology Mandi, Mandi-175005 (H.P.), India}


\author{Sumanta Tewari}
\affiliation{Department of Physics and Astronomy, Clemson University, Clemson, South Carolina 29634, USA}

\author{Tudor Stanescu}
\affiliation{Department of Physics and Astronomy, West Virginia University, Morgantown, West Virginia 26506, USA}


\begin{abstract}
Focusing on the implications of recent experiments on Majorana zero modes in semiconductor-superconductor (SM-SC) heterostructures, we critically examine the quantization of the zero-bias differential conductance as a possible unambiguous signature of Majorana physics in the presence of disorder. By numerically calculating the zero-bias conductance (ZBC) maps as function of Zeeman splitting and chemical potential for different disorder realizations, we find that the large topological region associated with the clean system, which is characterized by a quantized ZBC height $2e^2/h$, breaks up into progressively smaller `islands'' as the disorder strength increases. For strong disorder we show that the presence of small islands with ZBC value (approximately) equal to $2e^2/h$, which we refer to as ``quantized islands'', 
represents a unique signature of Majorana physics supporting  partially-separated Majorana modes (ps-MMs). Because of the small area/volume of these quantized islands in the parameter space, 
observing them in experiments may require sample selection and the systematic scanning of a large volume in the control parameter space. Upon decreasing disorder, the quantized islands increase in size and eventually coalesce into large topological regions. We conclude that the observation of quantized islands with ZBC value approximately equal to $2e^2/h$ demonstrates unambiguously the presence of the key ingredients necessary for Majorana physics, provides an excellent diagnostic tool for evaluating the disorder strength, and, consequently, represents the next natural milestone in the Majorana search.  

\end{abstract}

\maketitle

\section{Introduction}

Majorana zero modes (MZMs) -- the condensed matter avatars of the Majorana fermion \cite{Majorana1937} -- are not only potential building blocks for topological quantum computation (TQC)~\cite{Kitaev2003,Nayak2008}, but, in themselves,  objects of deep fundamental interest~\cite{wilczek1982quantum}. These  quasiparticles emerge in a class of condensed matter systems called topological superconductors and obey non-Abelian exchange statistics~\cite{moore1991nonabelions,read2000paired,nayak19962n}, a property with no correspondent in high energy physics. One of the simplest proposed realizations of MZMs involves a strong spin-orbit coupled semiconductor (SM) nanowire with proximity induced superconductivity (SC) and the presence of a magnetic field applied parallel to the wire~\cite{sau2010generic, sau2010non,oreg2010helical,lutchyn2010majorana}. The apparent simplicity of this proposal and its major practical and basic science implications  has spurred tremendous experimental activity in recent years~\cite{mourik2012signatures,deng2012anomalous,das2012zero,rokhinson2012fractional,churchill2013superconductor,finck2013anomalous,deng2016majorana,zhang2017ballistic,chen2017experimental,nichele2017scaling,albrecht2017transport,o2018hybridization,shen2018parity,sherman2017normal,vaitiekenas2018selective,albrecht2016exponential,Yu_2021}.

Despite remarkable progress, the unambiguous demonstration of topological MZMs  remains an outstanding experimental challenge. The most accessible experimental feature consistent with the presence of MZMs, the emergence of a zero-bias peak in differential tunneling conductance that is robust against variations of the magnetic field, has been observed by several groups in multiple samples~\cite{mourik2012signatures,deng2012anomalous,das2012zero,churchill2013superconductor,finck2013anomalous,deng2016majorana,zhang2017ballistic,chen2017experimental,nichele2017scaling,albrecht2017transport,o2018hybridization,shen2018parity,sherman2017normal,vaitiekenas2018selective,albrecht2016exponential,zhang2021,Yu_2021}. The Achilles' heel of this test is that certain topologically trivial states can also give rise to similar features~\cite{Brouwer2012,Mi2014,Bagrets2012,pikulin2012zero,ramon2012transport,  pan2020physical,moore2018two,moore2018quantized,vuik2018reproducing,Stanescu_Robust,added_Loss_2018prb_abs,ramon_Jorge2106exceptional,ramon2019nonhermitian,Jorge2019supercurrent, ramon2020from,  Jorge2021distinguishing}.
A more stringent condition is the quantization of the zero-bias conductance peak at zero temperature, which is predicted to give rise to quantized zero-bias conductance (ZBC) plateaus of height $2e^2/h$ as a function of control parameters such  as the magnetic field and gate potentials~\cite{sengupta2001midgap,law2009majorana, flensberg2010tunneling}. Recently reported observations of this type of feature~\cite{zhang2021} involve a high degree of sample selection and parameter fine tuning and  have generated a heated controversy regarding not only the nature of the underlying low-energy states, but key aspects of the experimental procedure. The current experimental situation is viewed as being affected by a ``confirmation bias'' problem~\cite{DasSarma2021} and a ``reproducibility crisis''~\cite{Frolov2021}. 
As shown in this study, the most likely current experimental situation involves strongly disordered nanostructures. Our  numerical calculations of the zero-bias conductance (ZBC) maps as function of Zeeman splitting and chemical potential show that, upon increasing the disorder strength, the large quantized topological region associated with the clean system breaks up  into progressively smaller quantized ``islands'' of ZBC value approximately equal to $2e^2/h$, while non-quantized islands with ZBC values exceeding $2e^2/h$ also emerge. 
In the strong disorder regime, because of the smallness of the islands, finding a quantized ZBC peak of height $2e^2/h$ is necessarily difficult and may require extensive tuning of the control parameters. Furthermore, we find that the mere observation of a quantized ZBC plateau as a function of a single control parameter does not provide sufficient evidence of Majorana physics, as it may be associated with the (quantized) boundary of a non-quantized island. However, as demonstrated below, the observation of finite quantized ``islands'' in a higher dimensional parameter space does represent an unambiguous signature of Majorana physics.    

Given the rather muddy status of the MZM search,  the key questions concern the path forward. We propose a paradigm for the MZM search based on three key elements: 1) {\em Basic assumption}: All experimentally-available Majorana hybrid nanostructures are affected by relatively strong parameter inhomogeneity, including random disorder. 2) 
{\em Basic tasks}: 
(i) Identify a meaningful operational definition of ``Majorana physics'' in the presence of disorder; 
(ii) Identify and characterize the main sources of disorder/inhomogeneity; 
(iii) Characterize in detail the low-energy physics of hybrid nanostructures in the presence of disorder/inhomogeneity. 3) {\em Basic approach}: Perform large-scale mappings of relevant observable quantities as functions of the control parameters. We note that basic task (i) is mainly theoretical and addresses a simple question: if a system contains all necessary Majorana ingredients (i.e., superconductivity, Zeeman splitting, and spin-orbit coupling) but disorder is strong enough to destroy the topological phase, can one meaningfully talk about Majorana physics? On the other hand, tasks (ii) and (iii) imply combined experimental and theoretical efforts involving different materials and device characteristics and observable features.

This paper is a theoretical contribution to this comprehensive program addressing tasks (i) and (iii) and focusing on the relevance of quantized ZBC peaks as a signature for Majorana physics in disordered systems. First, we show that, in the presence of disorder, remnant Majorana physics can be operationally understood as emerging locally within a certain disorder-controlled length scale and being characterized by the presence of partially-separated Majorana modes (ps-MMs). The ps-MMs require the presence of the key Majorana ``ingredients'' (i.e., superconductivity, spin-orbit coupling, and Zeeman splitting) and are adiabatically  connected to the topological MZMs upon formally expanding the length scale associated with the emergence of remnant Majorana physics. Next, we investigate the dependence of the zero-bias differential conductance on the Zeeman energy and the chemical potential in the presence of disorder by calculating numerically the corresponding two-dimensional maps. We find that the regions with quantized ZBC are either i) finite area islands or ii) boundaries of islands characterized by ZBC values that exceed $2e^2/h$. We show that the latter, non-quantized, islands are generated by Andreev bound states (ABSs) consisting of strongly overlapping pairs of Majorana modes, while the finite area quantized islands with ZBC values $\sim {2e^2}/{h}$ are always associated with the presence of ps-MMs and, therefore, represent a signature of remnant Majorana physics. The quantized islands can occur both inside and outside the nominally-topological region of the parameter space, while the non-quantized islands emerge outside this region. In the low-disorder limit, the non-quantized islands disappear, while the quantized islands coalesce into a large ``quantized continent'' within the topological region corresponding to the clean system. In the opposite, strong-disorder limit, the islands move outside the nominally-topological region, shrink, and may eventually disappear.

Our findings  suggest that the current experiments are conducted on strongly disordered nanostructures, which warrants the basic assumption of the paradigm proposed above. In this scenario, finding a quantized ZBC plateau as a function of a control parameter, e.g., the applied magnetic field, is necessarily hard and requires extensive fine tuning of all control parameters. We emphasize, however, that a quantized ZBC plateau as a function of a single control parameter does not provide clear evidence of Majorana physics, as this can be generated by a cut in parameter space tangent to the (quantized) boundary of an ABS-induced non-quantized island, as explicitly shown below (see, e.g., Fig.~\ref{FIG8}). By contrast, the observation of (even small) quantized islands would constitute strong evidence of remnant Majorana physics and partially-separated Majorana modes. In other words, such an observation would demonstrate that all necessary ingredients for realizing topological superconductivity and MZMs are actually present in SM-SC hybrid structures.  
Furthermore,  progress in materials growth and device engineering aimed at reducing disorder can be conveniently traced by systematically studying the number, size and position of quantized and non-quantized islands within large regions of the control parameter space. Since the basic ingredients necessary for realizing ps-MMs are the same as those required by topological MZMs, while the corresponding homogeneity conditions are significantly less restrictive, the unambiguous experimental demonstration of quantized ZBC islands  represents the next natural milestone in the Majorana search.  

\section{Remnant Majorana physics in disordered systems: operational approach} \label{approach}

The simplest effective model that describes the low energy physics of a semiconductor-superconductor (SM-SC) hybrid structure is given by the Bogoliubov-de Gennes (BdG) Hamiltonian
\begin{eqnarray}
H &=& -t\!\!\sum_{\langle i,j\rangle,\sigma}c_{i\sigma}^\dagger c_{j\sigma} + \sum_{i,\sigma}[V(x_i)-\mu] n_{i\sigma} +\Gamma\!\sum_{i}c_i^\dagger \sigma_x c_i 
\nonumber \\
&+& \frac{i\alpha}{2}\!\!\sum_{\langle i,j\rangle}\!\left(c_i^\dagger \sigma_y c_j \!+\! H.c.\right) +\Delta\sum_i\left(c_{i\uparrow}^\dagger c_{i\downarrow} \!+\! H.c.\right),  \label{Ham}
\end{eqnarray}
where $\langle i,j\rangle$ are nearest-neighbor sites on a a one-dimensional lattice, $c_i^\dagger = (c_{i\uparrow}^\dagger, c_{i\downarrow}^\dagger)$ is the electron creation operator on site $i$, $n_{i\sigma} = c_{i\sigma}^\dagger c_{i\sigma}$ is the number operator, and $\sigma_\nu$ (with $\nu=x, y, z$) are Pauli matrices. The model parameters $t$, $\mu$, $\Gamma$, $\alpha$, and $\Delta$ represent the nearest-neighbor hopping, chemical potential, Zeeman splitting, Rashba spin-orbit coupling, and induced pairing potential, respectively. In the presence of inhomogeneity and/or random disorder all these parameters are, in principle, position dependent. Here, we only consider potential disorder, which is incorporated through the position-dependent effective potential $V(x_i)$. We note that a more realistic modeling of hybrid SM-SC nanowires incorporates multi-band physics \cite{multiband}, proximity-induced energy renormalization effects \cite{renorm}, as well as various types of electrostatic effects \cite{electrostatics}. These additional ingredients are critical for developing an understanding of the quantitative impact of various materials and control parameters on the low-energy physics of the hybrid system. However, assuming that the semiconductor spectrum is characterized by an interband spacing larger than the other relevant energy scales, many of the effects generated by the realistic additional ingredients can be absorbed into renormalized effective parameters for Hamiltonian (\ref{Ham}). Most importantly, these effects do not modify qualitatively the basic mechanism for the emergence of topological superconductivity and MZMs predicted by the toy model. The interband spacing in the experimental system is given by the energy associated with the confinement in the transverse directions, but also on other effects such as the gate voltage, work function
difference between the semiconductor and the superconductor, and the
subband occupation. As discussed in Ref.~\cite{subBand_2020prb_woods}, the average interband separation for a wire of radius $35$ nm is $10-20$ meV,
and for a wire of radius $50$ nm is $5-10$ meV (see Figs. 5 and 6). These values are sufficiently bigger than the parameters considered in the present work, e.g., induced pairing potential $\sim 0.25$ meV, typical Zeeman field $\sim 1$ meV, spin orbit coupling $\sim 1.4$ (or $2.2$) meV for $\{\mu, \Gamma\}=\{2, 1.5\}$ (or \{6, 3\}) meV etc. justifying the 1D model.

\begin{figure}[t]
\begin{center}
\includegraphics[width=0.44\textwidth]{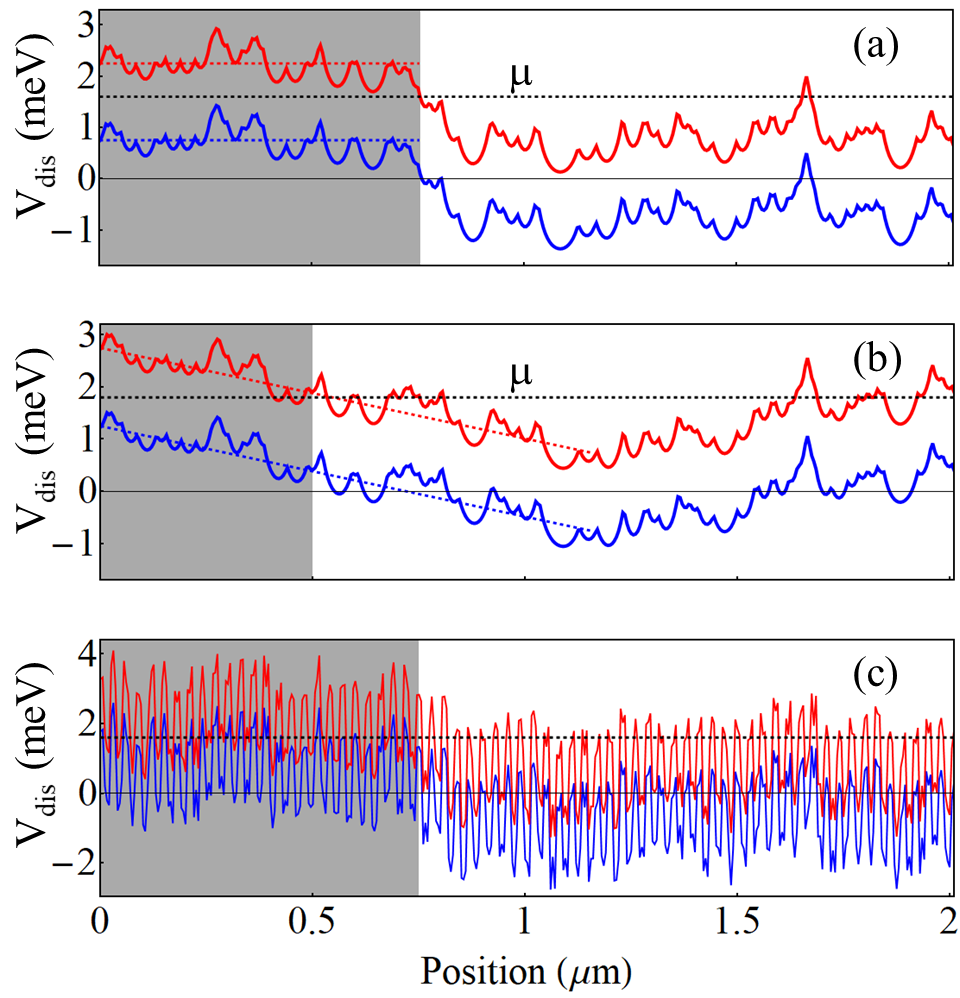}
\end{center}
\vspace{-3mm}
\caption{Position dependence of the Zeeman-split disorder potential, $V_{dis}^+(x)$ (red lines) and $V_{dis}^-(x)$ (blue lines), for a wire of length $L=2~\mu$m and a Zeeman field $\Gamma=0.75~$meV. The black dashed lines show the position of the chemical potential. In panels (a) and (b) the topological condition given by Eq. (\ref{topoC}) is manifestly satisfied within the shaded regions. In panel (c) the condition is effectively satisfied by the relevant, slowly varying component of the disorder potential (see the main text).}
\label{FIG1}
\vspace{-1mm}
\end{figure}

The key requirements for the emergence of MZMs are the presence of (induced) superconductivity (i.e., finite $\Delta$), spin-orbit coupling (finite $\alpha$) and Zeeman splitting (finite $\Gamma$), as well as the realization of the so-called topological condition, which for the model given by Eq. (\ref{Ham}) takes the form
\begin{equation}
[V^+(x_i) -\mu][\mu - V^-(x_i)] > \Delta^2, \label{topoC}  
\end{equation}
where $V^{\pm}(x_i) = V(x_i) \pm \Gamma$ is the Zeeman-split effective potential.  Intuitively, condition (\ref{topoC}) can be viewed as the requirement of having the chemical potential within the gap between $V^-$ and $V^+$ and no closer than $\Delta$ from either of the two limits. In principle, such a condition is always realized for large-enough values of $\Gamma$. In practice, however, the applied magnetic field  is limited by the collapse of the (parent) superconducting gap, hence $\Gamma$ cannot be arbitrarily large. Therefore, in a nonuniform system with a position-dependent  effective potential $V(x_i)$ it is possible that, for accessible values of $\Gamma$, condition (\ref{topoC}) is only realized locally, within a certain region of length $\delta L$ smaller than the length $L$ of the nanowire. If $\delta L$ is larger than a certain characteristic Majorana length scale, we operationally define the emergence of such a region as {\em remnant Majorana physics} and dub the corresponding near-zero energy modes as {\em partially-separated Majorana modes} (ps-MMs), also known as partially-separated Andreev bond states (ps-ABSs) \cite{moore2018two,moore2018quantized,czeng2019psABS_Kitaev,girish2020hybridization} or quasi-Majorana modes (q-MMs) \cite{vuik2018reproducing,czeng_qMs}. We note that the relevant Majorana length scale depends on nature of the ps-MMs \cite{Stanescu_Robust}. If the two Majorana modes are associated with the same spin subband, their overlap energy $\epsilon_M$ collapses toward zero if $\delta L > \xi_M$, where $\xi_M$ is the coherence length typically associated with the exponential decay of the Majorana wave function. By contrast, if the two Majorana modes are associated with opposite spin subbands (e.g., in the presence of an effective potential with nonzero average slope, or a step-like potential), the condition becomes $\delta L > \delta_M$, where $\delta_M$ is the inverse characteristic Fermi k-vector, or the width of the main peak of the Majorana wave function \cite{Stanescu_Robust}. In general, e.g., in the presence of disorder, the Majorana modes have mixed spin character and the relevant Majorana length scale can only be determined numerically. Nonetheless, the robust collapse toward zero of $\epsilon_M$ over finite ranges of control parameters is always associated with a spatial separation of the Majorana modes.

\begin{figure}[t]
\begin{center}
\includegraphics[width=0.49\textwidth]{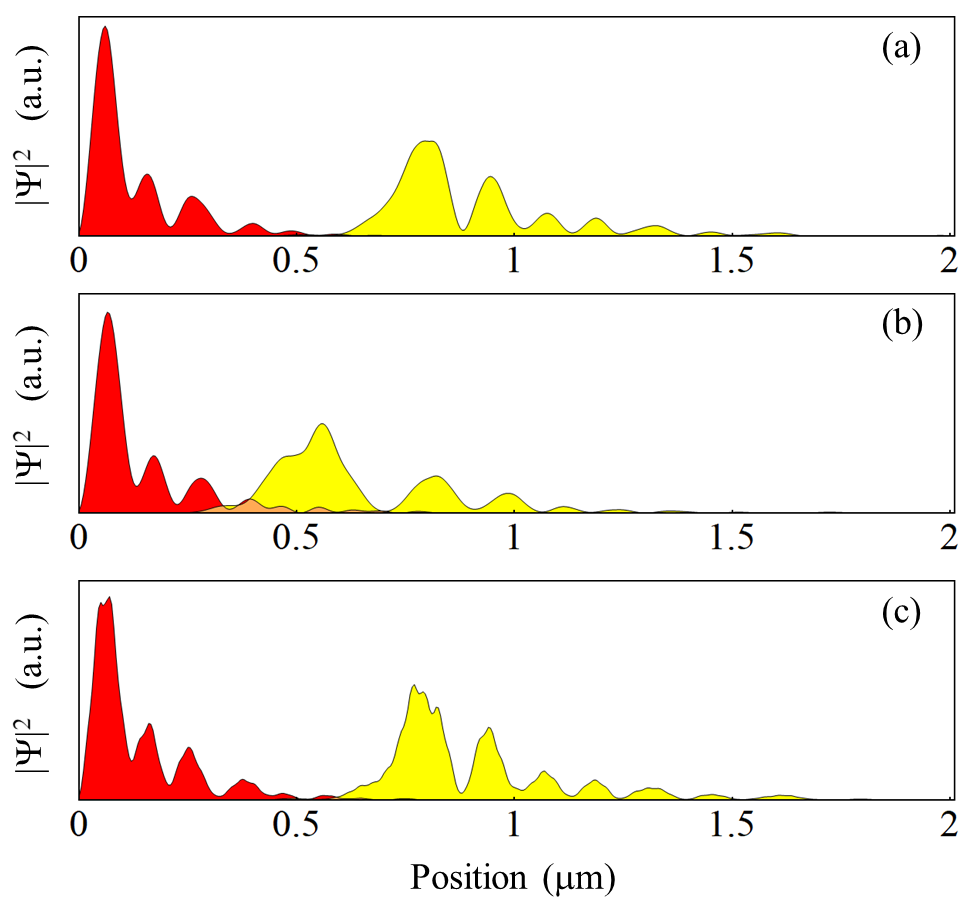}
\end{center}
\vspace{-3mm}
\caption{Majorana wave functions associated with the lowest BdG modes corresponding to the disorder realizations shown in Fig. \ref{FIG1}. Note that the topological condition given by Eq.~(\ref{topoC}) being (effectively) satisfied within the shaded regions in Fig. \ref{FIG1} translates into the emergence of pairs of ps-MMs at the edges of these regions.}
\label{FIG2}
\vspace{-1mm}
\end{figure}

For concreteness, and to gain further intuition, we consider three specific examples of position-dependent effective potentials corresponding to a system of length $L=2~\mu$m, as shown in Fig. \ref{FIG1}. We note that amplitude and characteristic length scale of the disorder potential depend on the source of disorder/inhomogeneity. For example, charge impurities generate an effective potential with amplitude of the order of several meV and relatively short characteristic length scales ($10-20~$nm) \cite{ChargeImp}. Disorder due to nonuniform oxidation of the surface of the parent superconducting film and inhomogeneities due to the presence of multiple voltage gates or due to strain in the nanostructure may involve larger length scales, although a detailed investigation of these sources of disorder/inhomogeneity remains a critical outstanding task. For the hybrid structures used in the laboratory it is likely that several different sources of disorder/inhomogeneity contribute simultaneously, generating effective potentials characterized by two or more length scales. In the illustrative examples shown in Fig.~\ref{FIG1} we have explicitly considered this possibility. Finally, we note that larger scale features can also be generated by sources with short characteristic length scales (e.g., by charge impurities) if they are non-uniformly distributed along the wire (e.g., if the impurity concentration varies significantly). 

The scenario shown in Fig.~\ref{FIG1}(a) corresponds to a set of parameters that satisfy the topological condition given by Eq.~(\ref{topoC}) within a segment of length $\delta L \approx 0.75~\mu$m near the left end of the wire (shaded area). Within this region, the disorder potential can be viewed as ``effectively weak'', and effective potential is equivalent to a weakly perturbed potential step~\cite{Stanescu_Robust}. Similarly, in Fig.~\ref{FIG1}(b) the effective potential near the left end of the wire can be viewed as a weakly perturbed smooth confinement potential~\cite{Stanescu_Robust}, and the topological condition is satisfied within the shaded segment of length $\delta L \approx 0.5~\mu$m. Note that in both cases the minimum Zeeman field required for satisfying the topological condition throughout the whole system is significantly higher than $\Gamma=0.75~$meV and may be practically inaccessible due to the collapse of the superconducting gap. 

The case shown in Fig.~\ref{FIG1}(c) is less straightforward. It is obvious that the topological condition (\ref{topoC}) is not explicitly  satisfied. However, the rapidly varying component of the effective potential has a weak effect on the low-energy physics. More specifically, we estimate the characteristic length scale of the low-energy BdG states $\psi_n$ localized within the shaded region as $\lambda\approx h/\sqrt{2m\Delta E} \approx 100~$nm, where $\Delta E$ is the average value of $V^+ -\mu$, and note that $\lambda$  is much larger than the period of the  rapidly oscillating component of $V_{dis}$.  Consequently,  if we write the disorder potential as the sum of fast and  slow varying components,  $V_{dis} = V_{dis}^{(fast)} + V_{dis}^{(slow)}$, we have $\langle \psi_n|  V_{dis}^{(fast)}|\psi_m\rangle \ll \langle \psi_n|  V_{dis}^{(slow)}|\psi_m\rangle$, which implies that $ V_{dis}^{(fast)}$ has negligible impact on the low-energy physics. Since $V_{dis}^{(slow)}$ is consistent with the the topological condition (\ref{topoC}), 
we conclude that the scenario in Fig.~\ref{FIG1}(c) is another example of remnant Majorana physics in the presence of disorder.

To further clarify this picture, we calculate the  Majorana wave functions associated with the lowest BdG mode corresponding to the disorder realizations illustrated in Fig.~\ref{FIG1}.  The results are shown in Fig. \ref{FIG2}. Note that the main peaks of the Majorana modes are separated by a distance $\delta L$ consistent with the length of the effectively topological (shaded) regions in Fig.~\ref{FIG1}. The emergence of these partially-separated Majorana modes is directly related to Eq.~(\ref{topoC}) being effectively satisfied locally. This is a clear signature of Majorana physics indicating (i) the presence of the key required ingredients (i.e., finite $\Gamma$, $\alpha$, and $\Delta$)   and (ii) the local (effective) realization of the topological condition. Furthermore, as a result of the spatial separation between the two Majorana modes (i.e., the red and yellow modes in Fig.~\ref{FIG2}), a tunnel probe at the left end of the wire will only couple to the leftmost (red) Majorana mode, which will lead to a quantized ZBC at zero temperature. This property is robust against (small) variations of the control parameters (e.g., Zeeman field and gate potentials). By contrast, any low-energy Andreev bound state that consists of nearly overlapping Majorana modes, which both couple to the tunneling probe, can only accidentally generate a quantized ZBC peak. Hence, remnant Majorana physics is necessarily associated with the emergence of ps-MMs and is manifested as robustness of certain properties (e.g., a quantized ZBC peak) against variations of the control parameters. 

Our next task is to explicitly test the generality of this framework. By focusing on one specific feature -- the quantized zero-bias differential conductance -- we (i) demonstrate that quantized ZBC ``islands'' emerging in a disordered system within a {\em finite volume} of a multi-dimensional parameter space are necessarily associated with the presence of MZMs or ps-MMs, hence represent a unique signature of (remnant) Majorana physics and (ii) investigate the evolution of the quantized islands as a function of disorder strength. Result (i) implies that  ZBC quantization is a good criterion for testing the presence of Majorana physics, but it should necessarily pass the robustness criterion within a multi-dimensional parameter space, i.e.,  one should demonstrate the presence of quantized islands, rather that quantized plateaus as function of a single parameter. Result (ii) provides a powerful tool for disorder diagnostics in hybrid nanostructures and demonstrates the critical importance of performing large-scale mappings of relevant observable quantities as functions of the control parameters, rather than focusing on fine tuning and data selection.

\section{Zero-bias conductance maps: disorder diagnostics and remnant Majorana physics} \label{ZBCmaps}

\begin{figure}[t]
\begin{center}
\includegraphics[width=0.45\textwidth]{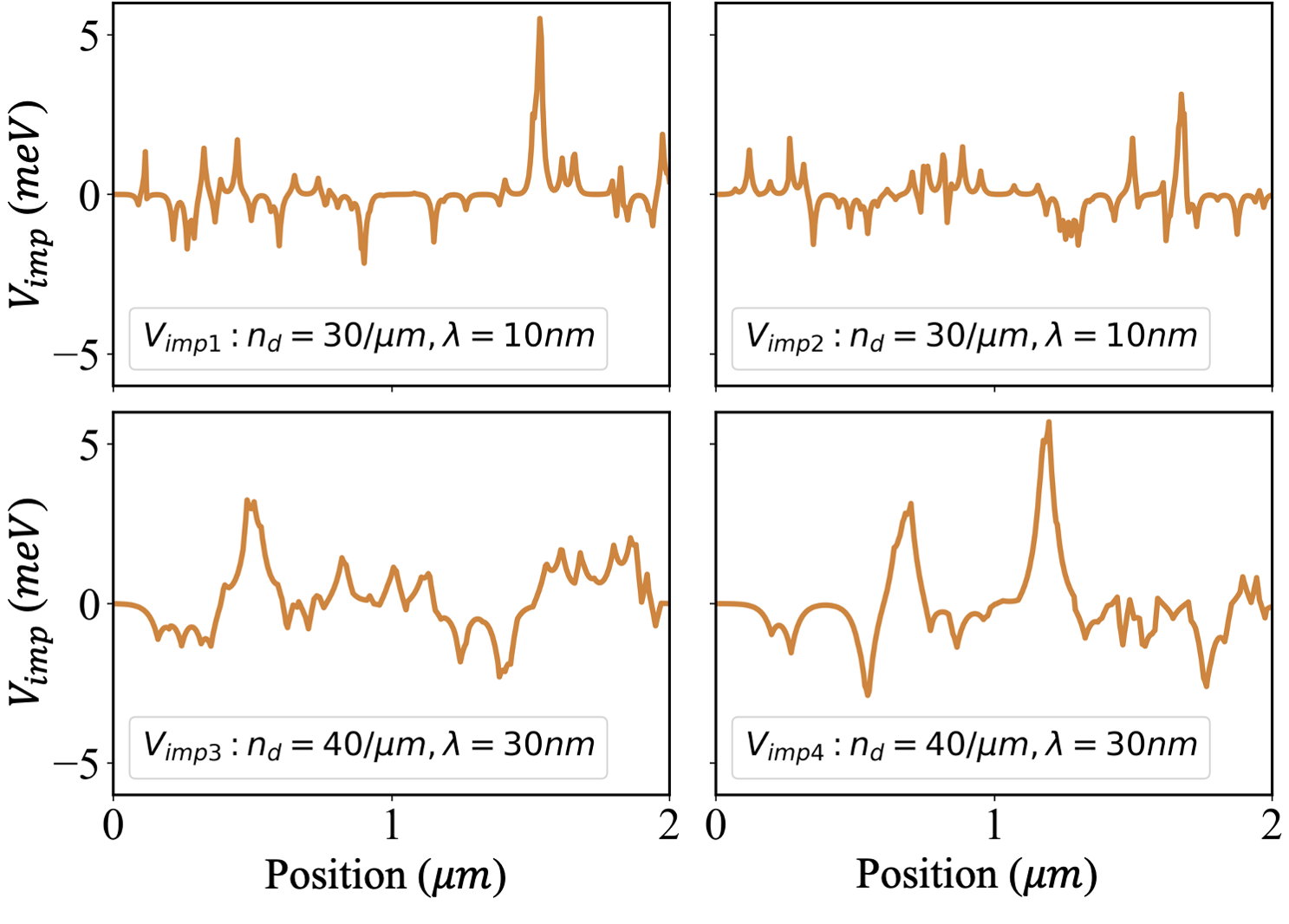}
\end{center}
\vspace{-3mm}
\caption{Spatial profiles  of four disorder potentials generated using Eq. \ref{Vdis} with $V_0=1$ and parameters $n_d$ and $\lambda$ given in the corresponding panels. Note that the positions $x_i$ of the impurities, as well as the corresponding amplitudes $A_i$ were randomly generated.}
\label{FIG3}
\vspace{-1mm}
\end{figure}

The analysis presented in this section is based on the numerical solution of the BdG equation corresponding to Hamiltonian (\ref{Ham}) for a system of length $L=2~\mu$m and different disorder potentials. The parameter values used in the calculation are: lattice constant $a=5~$nm, hopping $t=50.8~$meV (which corresponds to an effective mass $m_{eff} = 0.03 m_0$, with $m_0$ being the electron mass), Rashba coefficient $\alpha = 5~$meV (i.e., $250~$meV$\cdot$\AA), and induced pairing potential $\Delta = 0.3~$meV. The chemical potential ($\mu$) and Zemman field ($\Gamma$) are used as control parameters. 
We model the effective disorder potential $V(x) =V_{dis}(x)$ as the total potential generated by $N_d$ randomly distributed short-range impurities \cite{ChargeImp}. The potential of a single impurity located at position $x_i$ is 
\begin{equation}
V_{imp}^{(i)}(x) = A_i \exp\left(-\frac{|x-x_i|}{\lambda}\right),
\end{equation}
where $\lambda$ is the characteristic length of the impurity potential and $A_i$ is a random amplitude characterized by a Gaussian probability distribution ${\cal P}(A)$.  Given a value of the (linear)  impurity density $n_d$, a wire of length $L$ will contain $N_d = n_d L$ randomly distributed  impurities, which generate the (total) effective potential
\begin{equation}
V_{dis}(x) = V_0 \sum_{i=1}^{N_d} V_{imp}^{(i)}(x) = V_0 \sum_{i=1}^{N_d}A_i\exp\left(-\frac{|x-x_i|}{\lambda}\right),  \label{Vdis}
\end{equation}
where $V_0$, a ``global'' disorder amplitude that enables the control of the disorder strength. We consider four specific disorder realizations with spatial profiles shown in Fig. \ref{FIG3} and three different values of the global amplitude, $V_0=1, 2.5$, and $5$, which is equivalent to having 12 different simulated wires. 

\begin{figure}[t]
\begin{center}
\includegraphics[width=0.5\textwidth]{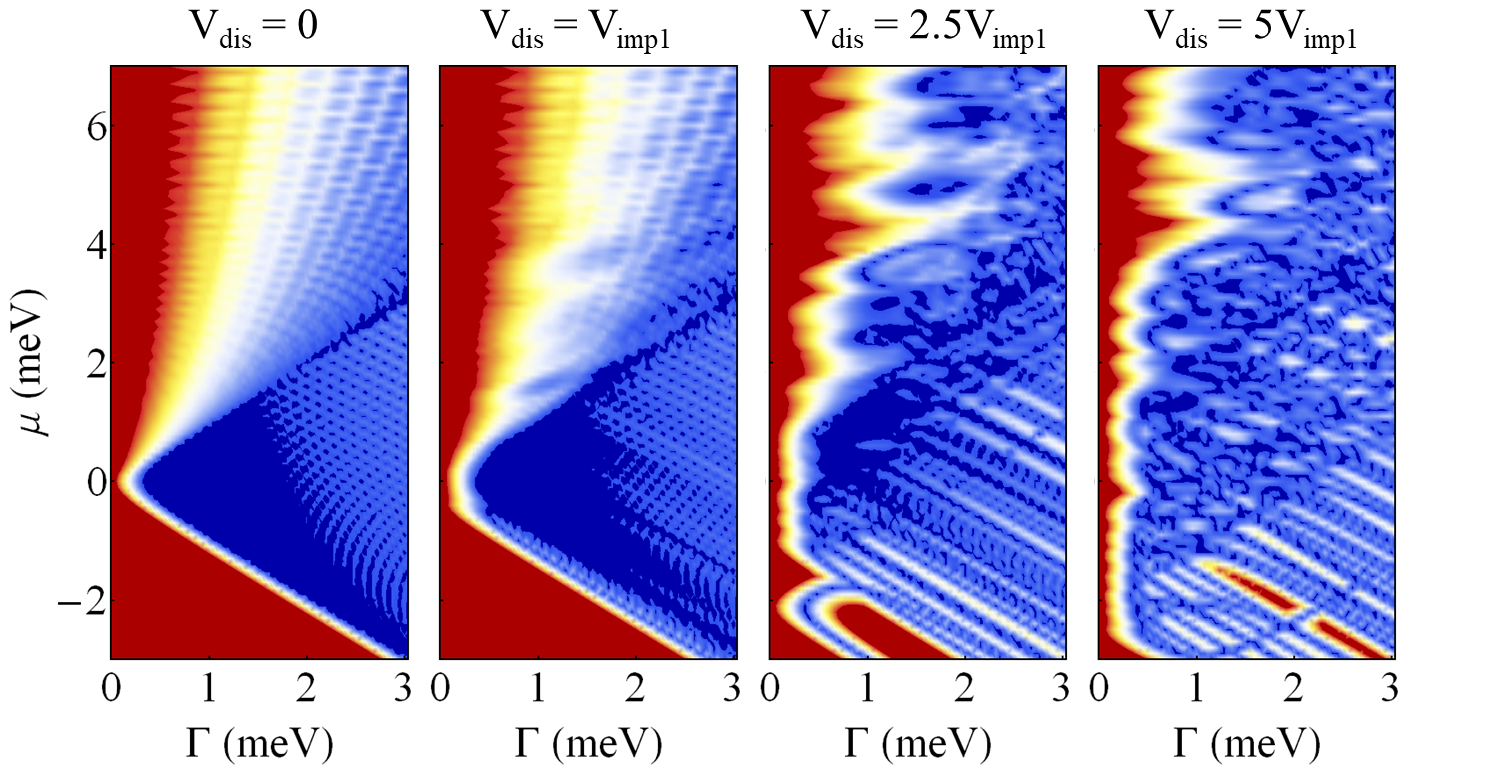}
\end{center}
\vspace{-3mm}
\caption{Energy of the lowest BdG state as a function of Zeeman splitting and chemical potential for a system with a disorder profile given by $V_{imp1}$  in Fig. \ref{FIG3} and different values of the global amplitude $V_0$. Dark blue indicates to the presence of near-zero energy states with $|E| < 10 ~\mu$eV. For $V_0=0$ (clean system), the low-energy states are Majorana modes emerging within the nominally-topological region $\Gamma^2 > \mu^2 + \Delta^2$. In a strongly disordered system, most of the low-energy states emerge outside the nominally-topological region within small islands in the parameter space.}
\label{FIG4}
\vspace{-1mm}
\end{figure}

The presence of disorder induces low-energy (sub-gap) states. To illustrate this point and to better understand the dependence of this property  on the control parameters, we calculate the ``phase diagrams'' corresponding to the energy of the lowest BdG state as a function of $\Gamma$ and $\mu$ for a system with $V_{dis} = V_0\times V_{imp1}$ and $V_0=0, 1, 2.5$, and $5$. The results are shown in Fig.~\ref{FIG4}. First, we note that the clean system ($V_{dis}=0$)  is characterized by the emergence of a large near-zero energy area (dark blue in Fig.~\ref{FIG4}) corresponding to the presence of Majorana zero modes within the nominally-topological region defined by the (clean) topological condition $\Gamma^2 > \mu^2 + \Delta^2$. Second, we note that Majorana physics is robust against weak disorder, which is reflected here by the relatively small changes in the phase diagram occurring as a result of having a nonzero disorder potential $V_{dis}=V_{imp1}$. However, upon increasing the amplitude of the disorder potential, the correlation between the area in parameter space characterized by low energy values and the (clean) topological phase diagram is completely lost. Also note that in the presence of strong disorder  most of the low-energy states emerge outside the nominally-topological region within small islands in the parameter space. The typical area of such an island decreases with increasing $V_0$. We emphasize that low-energy states responsible for the emergence of these islands are generally not localized near the ends of the wire and, consequently, remain ``invisible'' to local probes connected to the edges of the system, such as, for example,  charge tunneling. If, however, a low-energy state is localized near the end of the wire, it generates a zero-bias conductance (ZBC) peak in a tunneling measurement. If the low-energy states are associated with a finite area low-energy island, the ZBC peak is robust against variations of the control parameters within a certain range given by the size of the island. Nonetheless, since the emerging low-energy state can be a MZM, or a ps-MM, or simply an Andreev bound state consisting of overlapping Majorana modes \cite{moore2018two}, the observation of robust ZBC peaks by itself does not demonstrate the presence of Majorana physics, i.e., the presence of MZMs or ps-MMs.  

Since a robust ZBC peak does not provide a selective-enough Majorana signature, we consider three additional criteria: a) the quantization of the ZBC peak in the zero temperature limit, b) the correlation between the area in the parameter space characterized by the emergence of ZBC and the topological phase diagram, and c) the presence of edge-to-edge Majorana correlations \cite{moore2018two}.  Criterion (c) is the most restrictive and is only consistent with the presence of MZMs localized at the two ends of the system. While this represents an excellent MZM signature, it requires (effectively) weak disorder, a condition that is probably not satisfied by currently available hybrid structures. Criterion (b) is consistent with the presence of either MZMs or ps-MMs and, in some cases, is less stringent regarding the acceptable strength of the disorder potential. For example, a system characterized by the disorder potential shown in Fig. \ref{FIG1}(a) is consistent with the emergence of ZBC (when tunneling into the left end of the system) inside a parameter region roughly corresponding to the topological phase of a clean wire (up to an overall shift of the chemical potential).  This is the case because the left (shaded) segment of the wire has effectively weak disorder. Note, however, that the system will not satisfy criterion (c) because the ``yellow'' Majorana has no significant weight at the right end of the wire, until much higher values of the Zeeman field, a regime that may be experimentally inaccessible because of the collapse of the superconducting gap. Analyzing the maps shown in Fig.~\ref{FIG4} reveals that the correlation between the low-energy areas and the nominally-topological region holds only for weak disorder ($V_{dis} = V_{imp1}$), being completely lost even for the intermediate case $V_{dis} =2.5 V_{imp1}$. 

We now turn our attention to criterion (a) -- the quantization of the ZBC peak -- and address the following key questions: (i) How does it compare with criterion (b) with respect to the disorder strength requirements. (ii) Is it selective enough, i.e., does it exclude the possibility of having garden variety ABSs as the source of the characteristic  observable feature? The first question can be reformulated in terms of the existence of remnant Majorana physics and ps-MMs -- which necessarily generate quantized ZBC peaks when one of the modes is localized near the end of the wire and is separated-enough from the other mode -- at disorder strengths higher than those consistent with criterion (b), when the correlation between the low-energy regions and the topological phase diagram has disappeared. The answer to this question is straightforward and can be obtained by simply inspecting the nature of the low-energy states in Fig. \ref{FIG4} for $V_0=2.5$ and $V_0=5$. Indeed, we find that some of these states are ps-MMs, which implies that Majorana physics survives at disorder strengths higher than those consistent with criterion (b). 

Previous theoretical studies have found that ZBC peaks generating quantized ZBC plateaus of height $2e^2/h$  as a function of control parameters such as the Zeeman field can be associated with genuine (topological) MZMs, ps-MMs \cite{moore2018quantized}, or even trivial (non-separated) ABSs \cite{pan2020physical}. This would suggest that  criterion (a) has no more selective power than the simple observation of (robust) ZBC peaks. Here, we strengthen the robustness requirements associated with this criterion and show explicitly that the observation of {\em quantized islands} having a finite volume in the parameter space is consistent with the presence of MZMs and ps-MMs, but cannot be associated  with the presence of trivial ABSs.  Consequently, we conclude that \textit{the observation of quantized islands represents an unambiguous signature of (remnant) Majorana physics.}  
 
\begin{figure}[t]
\begin{center}
\includegraphics[width=0.46\textwidth]{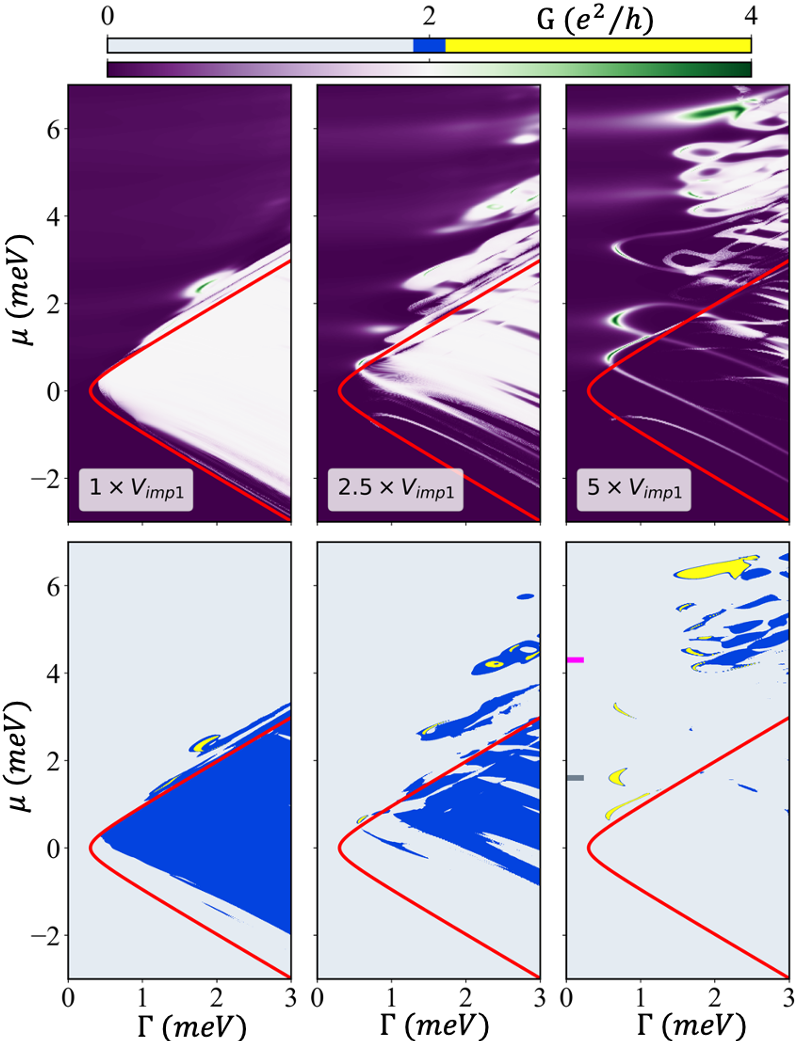}
\end{center}
\vspace{-3mm}
\caption{{\em Top panels}: Zero-bias conductance maps for a  disordered system with $V_{dis}=V_0\times V_{imp1}$.  {\em Bottom panels}: Three-color ZBC maps. Blue corresponds to conductance values within a $\pm 5$\% window of the quantized conductance, $2 e^2/h$, while  yellow and gray correspond to conductance values above and below this window, respectively. The topological phase boundary of the clean system is marked by a red line. The system with low disorder ($V_0=1$) is characterized by a large quantized area (blue) that almost coincides with topological region. Upon increasing the disorder strength, this area breaks into small islands that migrate outside the nominally-topological region and are not necessarily quantized.}
\label{FIG5}
\vspace{-1mm}
\end{figure}

To properly characterize the presence of quantized ZBC islands, we calculate the ZBC maps as function of the Zeeman field and chemical potential. The results corresponding to a disordered system with $V_{dis}=V_0\times V_{imp1}$ are shown in Fig.~\ref{FIG5}. The topological phase boundary corresponding to a clean system is marked by a red line.
First, note that for weak disorder ($V_0=1$) the high conductance area almost coincides with the nominally topological region. Upon increasing the disorder strength, this large area breaks into smaller islands located both inside an outside the topological region. Some of these islands are quantized (blue islands in Fig. \ref{FIG5}), while others are characterized by ZBC values larger than $2e^2/h$. The typical area of an island reduces with increasing disorder strength and the location of the islands changes from inside to outside of the nominally-topological region. To understand the ``migration’’ of quantized islands outside of the nominally-topological region, we first note that the average over position of the disorder potential is zero, $\langle V_{dis}\rangle = 0$, which means that the ``migration’’ is not the result of an overall disorder-induced shift of the chemical potential. However, the relevant low-energy states $|\psi\rangle$ do not extend uniformly throughout the whole wire, so that the quantity relevant to the ``Majorana condition’’ in the presence of disorder is not the (nominal) chemical potential $\mu$, but rather the ``effective’’ chemical potential $\mu_{eff} = \mu + \langle\psi|V_{dis}|\psi\rangle$.  Consequently, one can have $\mu_{eff}$ within the nominally-topological region (i.e., satisfying the ``Majorana condition’’), while $\mu$ lies outside the topological boundary predicted by the disorder-free model. In addition, we note that the states corresponding to low values of the chemical potential tend to be very localized, typically away from the ends of the wire, and, consequently, do not contribute to the zero-bias differential conductance. By contrast, the states corresponding to large values of $\mu$ are more delocalized and may couple strongly to the lead, generating signatures in the ZBC maps.

\begin{figure}[t]
\begin{center}
\includegraphics[width=0.46\textwidth]{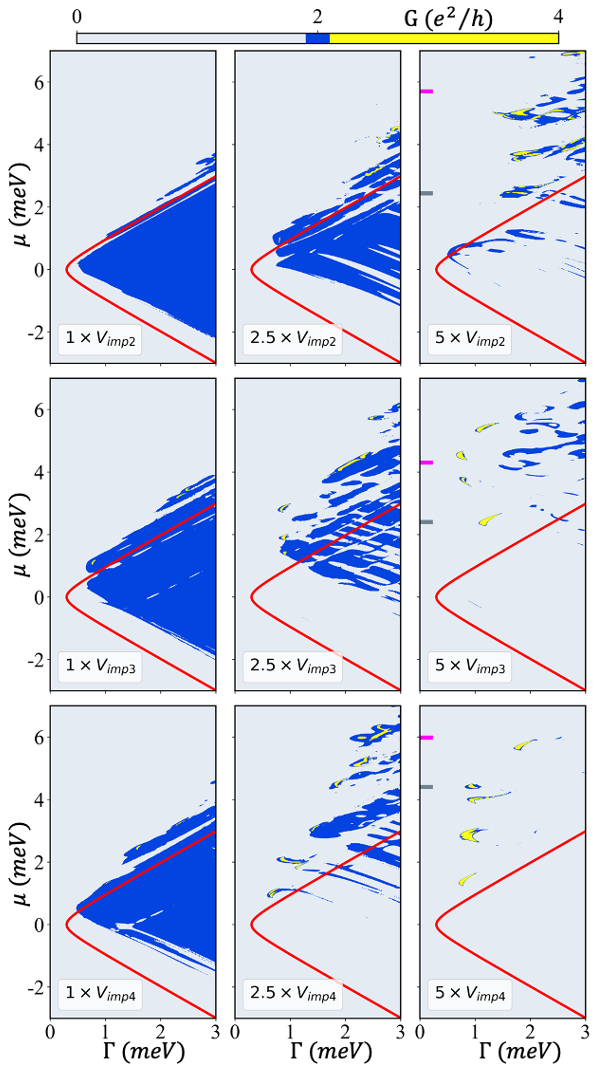}
\end{center}
\vspace{-3mm}
\caption{Zero-bias conductance maps for a disordered system with $V_{dis}=V_0\times V_{imp2}$ (top panels), $V_{dis}=V_0\times V_{imp3}$ (middle panels), and $V_{dis}=V_0\times V_{imp4}$ (bottom panels). Blue corresponds to conductance values within a $\pm 5$\% window of the quantized conductance, $2 e^2/h$, while  yellow and gray correspond to conductance values above and below this window, respectively. The short magenta and gray lines inside the right panels indicate the locations of the constant chemical potential cuts  shown in Fig. \ref{FIG7}.}
\label{FIG6}
\vspace{-1mm}
\end{figure}

To verify the generality of these trends, we calculate the ZBC maps corresponding to the other three disorder profiles from Fig.  \ref{FIG3}. The results shown in Fig. \ref{FIG6} clearly confirm the trends discussed above and reveal the quantitative differences between different disorder realizations. Focusing on the stronger disorder cases ($V_0=5$), we note that the large conductance areas consist of relatively small quantized  (blue) and non-quantized (yellow) islands typically located outside the nominally-topological region. Of course, the boundary of a yellow island is always quantized, i.e., blue, as the ZBC varies continuously from values larger that $2e^2/h$ inside the island to values below $2e^2/h$ outside the island. We also note that the map corresponding to $V_{dis}=5\times V_{imp4}$ has only small non-quantized (yellow) islands scattered outside the nominally-topological region. This property is also associated with the other disorder profiles, but occurs at larger values of the global disorder amplitude $V_0$. Hence, we conclude that in the strong disorder limit finding quantized ZBC peaks is difficult and may require a lot of fine tuning and sample selection. In particular,  if quantized (blue) islands are absent, the only way of obtaining a quantized ZBC peak is by tuning the system near the (quantized) boundary of a yellow island. If accidentally this boundary is relatively straight along the direction of the driving parameter, this may result in a quantized ZBC plateau.

\begin{figure}[t]
\begin{center}
\includegraphics[width=0.46\textwidth]{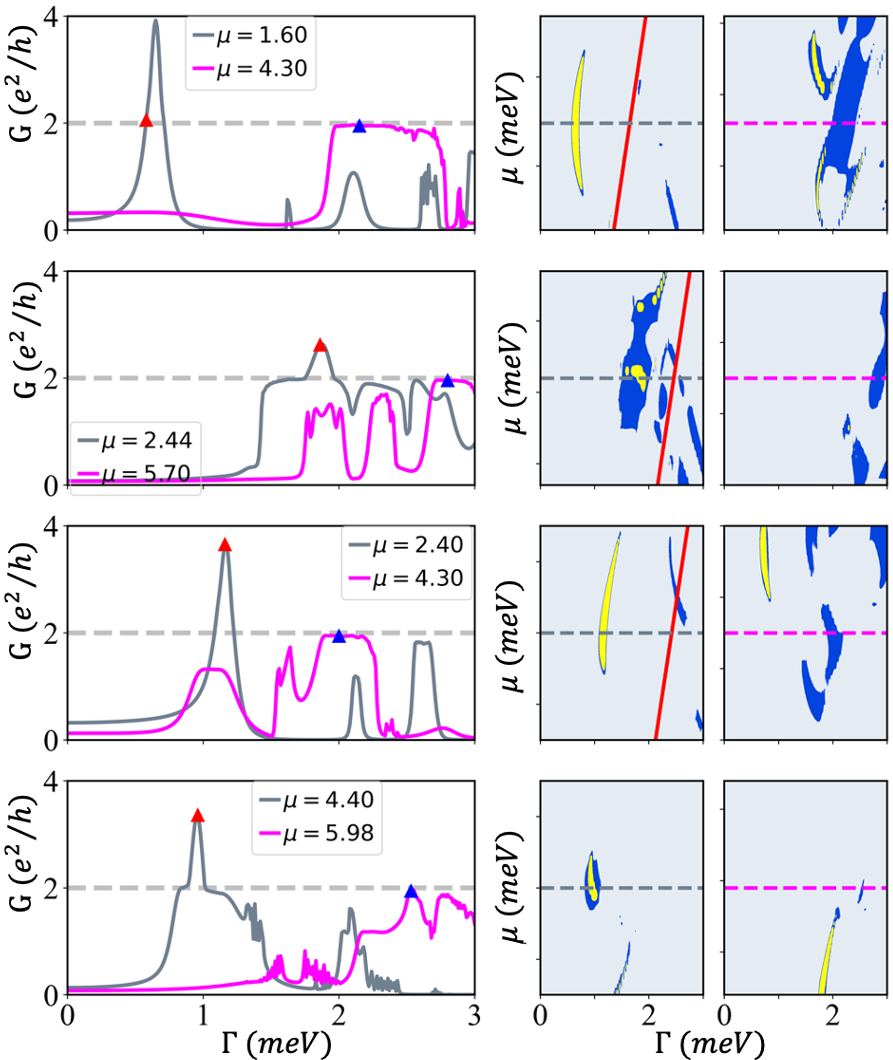}
\end{center}
\vspace{-3mm}
\caption{Constant chemical potential traces through representative quantized and non-quantized islands. The corresponding values of the chemical potential are marked  by short magenta and gray lines in the right panels  of Figs. \ref{FIG5} and \ref{FIG6}. The panels correspond, from top to bottom, to $V_{dis} = 5\times V_{imp1}, ~5\times V_{imp2}, ~5\times V_{imp3}, ~{\rm and}~ 5\times V_{imp4}$, respectively. Note that cutting through a finite quantized (blue) area translates into a (finite width) ZBC plateau as a function of the Zeeman field.}
\label{FIG7}
\vspace{-1mm}
\end{figure}

To further characterize the high-conductance islands, we calculate constant chemical potential ZBC traces through representative quantized (blue) and non-quantized (yellow) islands. The corresponding values of the chemical potential are marked  by short magenta and gray lines in the right panels  of Figs. \ref{FIG5} and \ref{FIG6}, while the corresponding traces are shown in Fig. \ref{FIG7}. Note that a well-defined quantized ZBC plateau is always associated with the presence of a finite area quantized (blue) islands. We also note that some of the quantized (blue) islands contain non-quantized (yellow) ``mountains''.  We still classify these ``mixed'' cases as ``quantized islands'', as long as they contain two-dimensional, finite area quantized  regions. This is in contrast to the purely yellow, non-quantized islands, which only have one-dimensional quantized boundaries.   
For example, the gray line in the bottom panel of Fig.~\ref{FIG7} corresponds to a ``mixed'' case. Notice, the narrow quantized plateaus on both sides of the ZBC maximum. Such plateaus are absent  in traces associated with the non-quantized (yellow) islands (see, e.g., the gray line fronm the top panel in Fig. \ref{FIG7}). 

\begin{figure}[t]
\begin{center}
\includegraphics[width=0.46\textwidth]{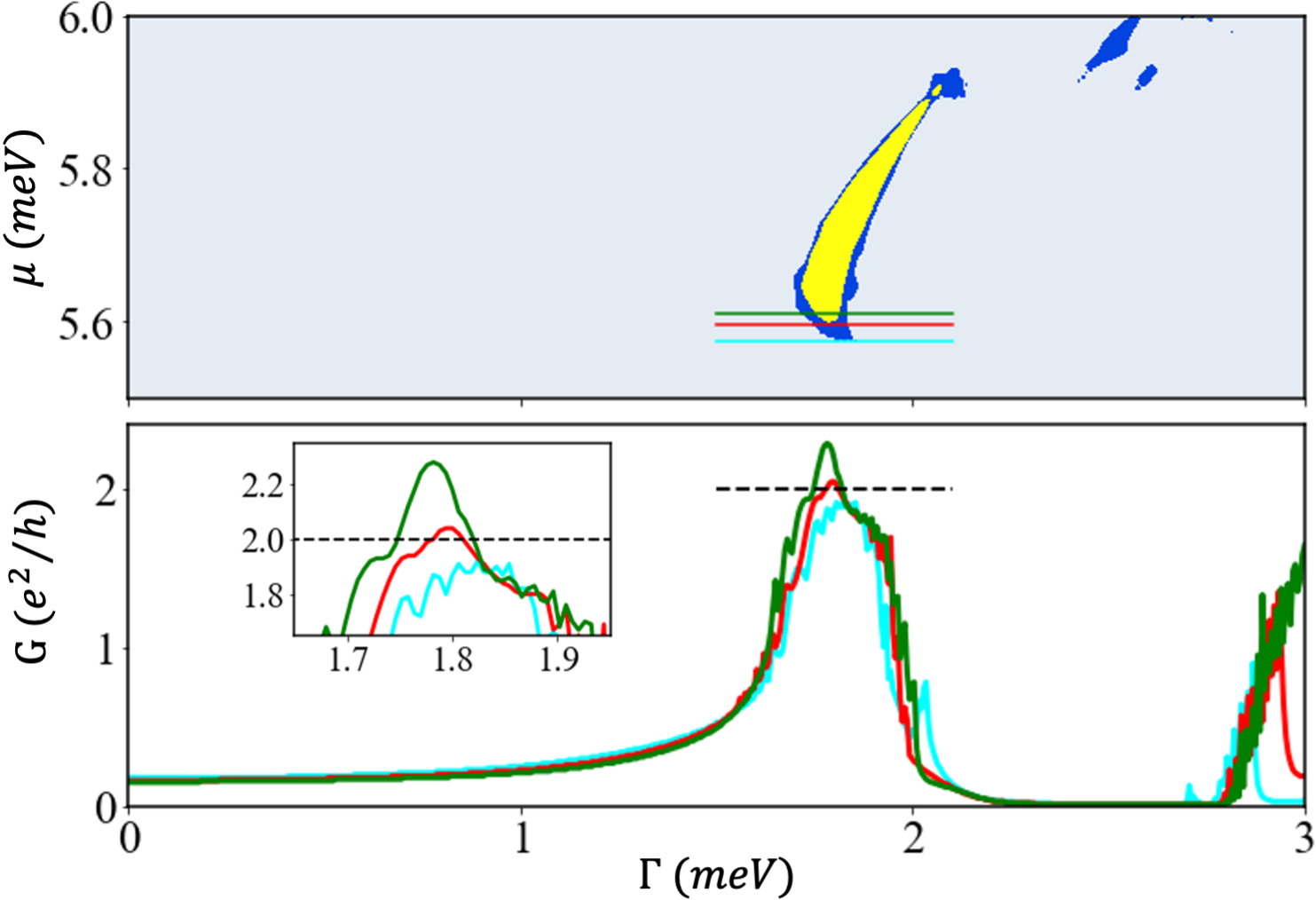}
\end{center}
\vspace{-3mm}
\caption{Traces obtained by fine tuning the chemical potential near the boundary of a non-quantized island corresponding to $V_{dis}=5\times V_{imp4}$. Note that small variations of the chemical potential lead to the disappearance of the quantized ``plateau''. The corresponding values of the chemical potential are: $\mu=5.58~$meV (cyan line), $\mu=5.60~$meV (red), and $\mu=5.61~$meV (green).}
\label{FIG8}
\vspace{-1mm}
\end{figure}

We emphasize the any quantized plateau that corresponds to a cut through the bulk of a quantized (blue) island is robust against small variations of the chemical potential. This is the very essence of having a ``quantized island'' of finite area/volume in the parameter space and an illustration of the enhanced robustness requirements that we apply to the quantized ZBC criterion. 
To contrast the robust ZBC plateaus associated with quantized islands with the quantized conductance peak obtained by fine tuning the control parameters near the boundary of a non-quantized island, we consider the  three cuts shown in Fig. \ref{FIG8}, which correspond to slightly different values of the chemical potential. Note that the red curve exhibits a (nearly) quantized maximum, but very small variations of the chemical potential destroy this quantization. This property is a direct consequence of the fact that robust quantization necessarily involves a finite area/volume in the parameter space and the boundaries of a non-quantized (yellow) island do not satisfy this requirement.  

\begin{figure}[t]
\begin{center}
\includegraphics[width=0.46\textwidth]{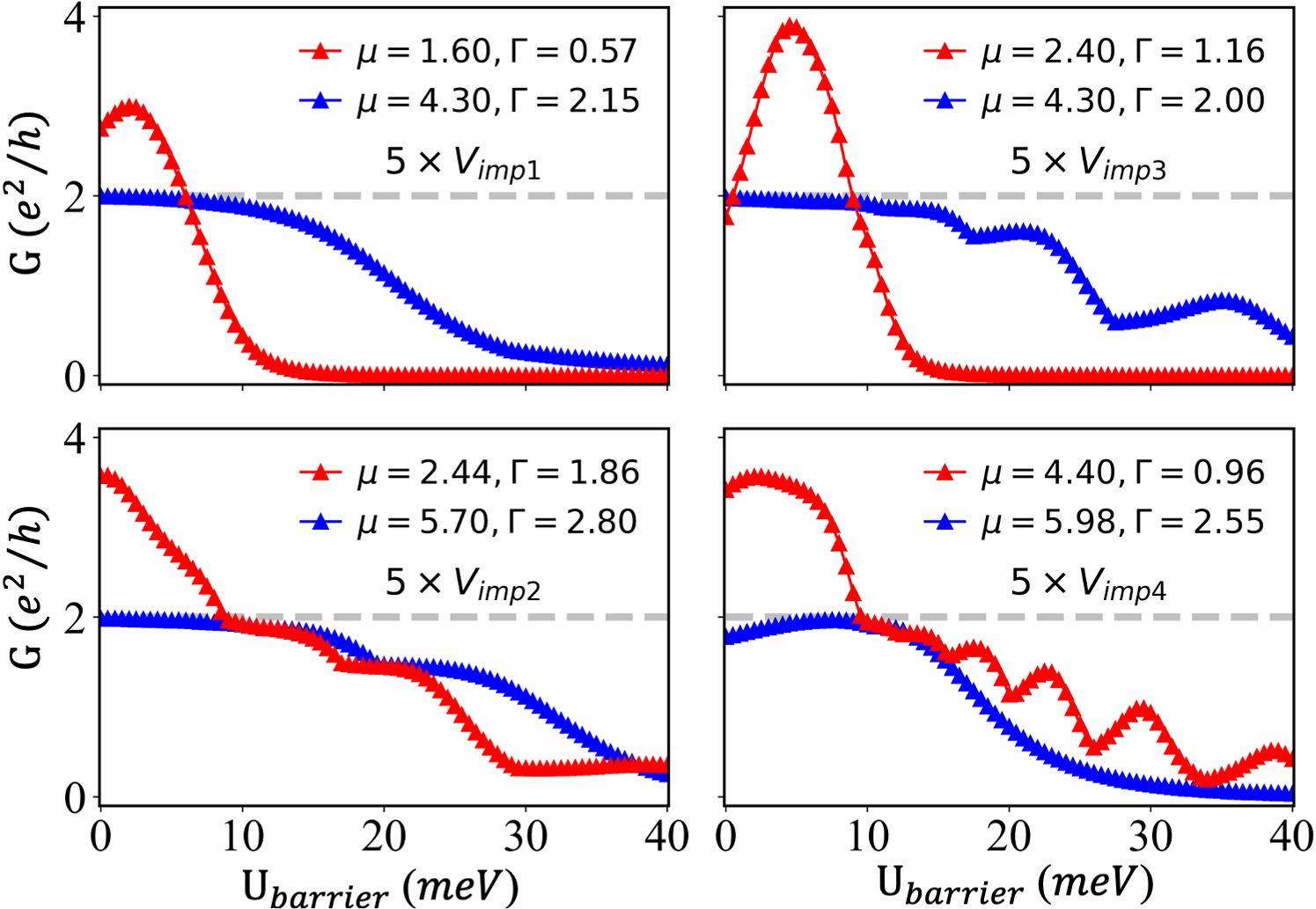}
\end{center}
\vspace{-3mm}
\caption{Dependence of the ZBC on the tunnel barrier height for parameters corresponding to the red and blue triangles in Fig. \ref{FIG7} (left panels). Blues triangles, which correspond to quantized islands, show quantized plateaus as a function of $U_{barrier}$. Narrower plateau-like features are also associated with ``mixed'' cases -- see the red curves in the lower panels. By contrast, the red curves in the upper panels are associated with non-quantized (yellow) islands and show no evidence of quantized plateaus. }
\label{FIG9}
\vspace{-1mm}
\end{figure}

Additional evidence regarding the robust quantization associated with the blue islands can be obtained by studying the dependence of the ZBC on the third available control parameter,  the height $U_{barrier}$ of the tunnel barrier. For concreteness, we focus on the parameters marked by red and blue triangles in the left panels of Fig. \ref{FIG7}. The results are shown in Fig. \ref{FIG9}. First, we notice that the curves marked by blue triangles, which correspond to parameters associated with quantized islands, exhibit a quantized ZBC plateau at small enough values of the barrier height. Even the  tiny blue island corresponding to $V_{dis}=5\times V_{imp4}$ near $\mu=5.98~$meV and $\Gamma=2.55~$meV has an almost quantized height over a significant range of $U_{barrier}$ values. Of course, upon increasing the barrier height, the ZBC peak becomes very narrow, as a result of reducing the effective coupling to the normal lead, and the quantized conductance value cannot be reached due to the presence of a finite broadening, $\eta = 0.4~\mu$eV. We have introduced a small finite broadening in the calculation to avoid the emergence of extremely narrow ZBC peaks, which, in practice, can never be observed. 
By contrast to the blue triangles, the red triangles corresponding to  $V_{dis}=5\times V_{imp1}$ and $V_{dis}=5\times V_{imp3}$, which are associated with non-quantized (yellow) islands exhibit no quantized plateau and generate ZBC values higher than $2e^2/h$ for certain values of the tunnel barrier height. Finally, the red triangles in the $V_{dis}=5\times V_{imp2}$ and $V_{dis}=5\times V_{imp4}$ panels correspond to mixed cases and are characterized by ZBC values that exceed $2e^2/h$ in the low $U_{barrier}$ limit, but exhibit narrow plateau-like features for $U_{barrier} \approx 10-13~$meV, indicating that the area corresponding to the yellow, non-quantized ``mountain'' inside the blue island decreases with increasing $U_{barrier}$.
The bottom line is that large quantized (blue) islands exhibit ZBC quantization that is quite robust against variations of the tunnel barrier height, small quantized islands are less robust against such variations, while non-quantized (yellow) islands generate ZBC quantization only accidentally, at very specific values of the control parameters. 

\begin{figure}[t]
\begin{center}
\includegraphics[width=0.46\textwidth]{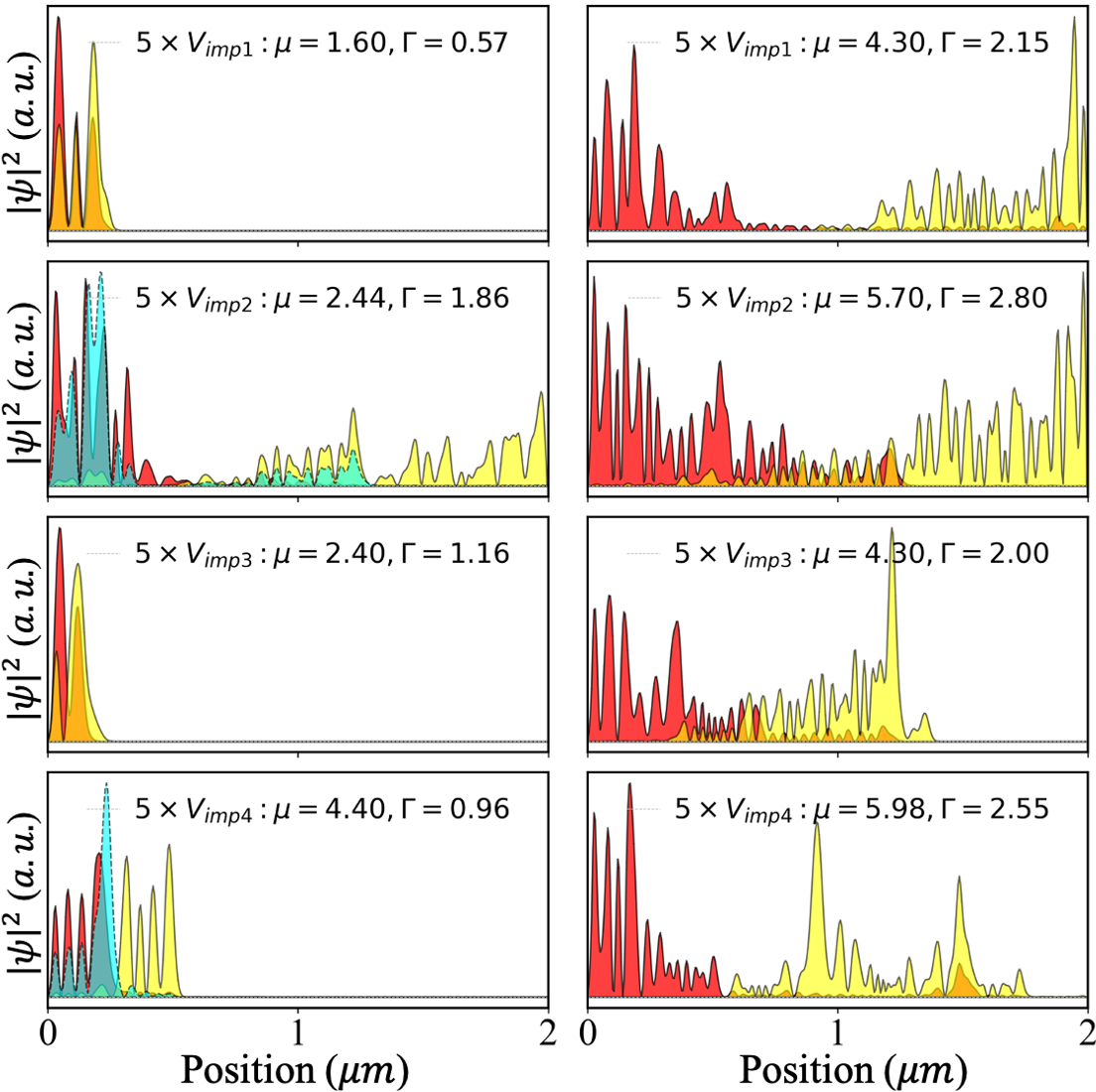}
\end{center}
\vspace{-3mm}
\caption{Majorana wave functions of the lowest energy BdG states (yellow and red) and leftmost Majorana modes of the third lowest states (cyan) for parameters corresponding to the triangles in Fig.~\ref{FIG7} (left panels). Note that for $5\times V_{imp2}$ and $5\times V_{imp4}$ the first three BdG modes are nearly degenerate. No other Majorana mode associated with low-energy BdG states overlaps with the leftmost red Majoranas. 
The left and right panels correspond to red (non-quantized ZBC) and blue (quantized ZBC) triangles, respectively. Note that, the Majorana wave functions associated to the blue triangles (right panels), which correspond to quantized islands, consist of partially-separated Majorana modes (ps-MMs). By contrast, the wave functions corresponding to red triangles (i.e., non-quantized ZBC) form pairs of highly overlapping Majorana modes (red-yellow or red-cyan), which generate regular ABSs.}
\label{FIG10}
\vspace{-1mm}
\end{figure}

The final element of our analysis consists of determining the nature of the low-energy states associated with different types of high-ZBC islands. In Fig. \ref{FIG10} we show the position dependence of the Majorana wave functions of the lowest energy BdG state corresponding to a few representative parameter sets associated with quantized (blue) and non-quantized (yellow) islands. The clearly distinguishing feature is that all low-energy states associated with quantized blue islands consists of partially-separated Majorana modes (ps-MMs), while the states associated with non-quantized yellow islands are regular ABS states consisting of pairs of  highly overlapping Majorana modes. In the mixed cases, the yellow ``mountains'' emerge because the leftmost (red) Majorana mode strongly overlaps with a Majorana mode corresponding to the third lowest BdG state (cyan in Fig. \ref{FIG10}), again forming a regular ABS localized near the left end of the system. 
Note that the robustness of the ZBC quantization is intrinsically connected to the spatial separation of the ps-MMs and is the result of i) having only one Majorana mode effectively coupled to the tunnel probe and ii) having the spatial separation (and, implicitly, the couplings of the Majorana modes to the probe) vary continuously with the control parameters.      
Since the emergence of ps-MMs  is directly connected to the presence of (remnant) Majorana physics, as discussed previously,
we conclude that the observation of quantized islands having a finite area/volume in the parameter space represents a unique signature of Majorana physics that cannot be mimicked  by garden variety low-energy Andreev bound states, which are ubiquitous in disordered systems.  

\section{Discussion and Conclusion}

We have investigated the quantization of the zero-bias differential conductance as a possible signature of Majorana physics in disordered semiconductor-superconductor hybrid structures.  By numerically calculating the zero-bias conductance (ZBC) maps as a function of Zeeman splitting and chemical potential, we have shown that, upon increasing the disorder strength, the large topological region associated with the clean system breaks up into progressively smaller quantized islands of ZBC value approximately equal to $2e^2/h$, simultaneously with the emergence of non-quantized islands characterized by ZBC values larger that $2e^2/h$. In the strong disorder regime, because of the smallness of the islands, finding a quantized ZBC peak of height approximately equal to $2e^2/h$ should be experimentally difficult and may require sample selection and extensive tuning of the control parameters. 
Nonetheless, we demonstrate that the very presence of quantized islands characterized by a ZBC value approximately equal to $2e^2/h$  and having a finite area/volume in a multi-dimensional parameter space is uniquely linked to the emergence of Majorana physics supporting Majorana zero modes (MZMs) or partially-separated Majorana modes (ps-MMs). By contrast, non-quantized islands characterized by ZBC values that exceed $2e^2/h$ are associated with the presence of trivial near-zero energy Andreev bound states that consist of highly overlapping Majorana components. 

In the strong disorder limit, we find that the high-conductance islands are typically small and located outside the nominally-topological region associated with the clean system. In this limit, the conductance maps may contain no quantized island. If this is the case, quantized ZBC peaks occur only accidentally at the (quantized) boundaries of non-quantized islands and observing them requires fine tuning the control parameters. Consequently, observing a quantized ZBC plateau as a function of a single control parameter is not a unique signature of Majorana physics,  as it can be obtained by moving along a (locally) straight-enough boundary of a non-quantized island , which is generated by a trivial low-energy  Andreev bound state (see Fig. \ref{FIG8}). By contrast, observing quantized islands of finite area/volume in parameter space represents an unambiguous demonstration of Majorana physics. Note, however, that the presence of ps-MMs (or even MZMs) does not automatically imply a quantized ZBC peak, e.g., if none of the Majorana modes is close enough to the end of the wire to ensure a measurable coupling to the tunnel probe. In other words, the quantized island Majorana criterion does not exclude false negatives. We also note that, although we have introduced the quantized island criterion using a specific model and a specific type of disorder, we expect our main conclusions to hold in general. This includes different types of quasi one-dimensional systems, such as, for example, nanowires realized in patterned two-dimensional electron systems hosted by semiconductor-superconductor heterostructures \cite{2016_2DEG_nc,2017_2DEG_prl} and full-shell nanowires \cite{2020full_shell,2021full_shell}, as well as different types of disorder/inhomogeneity, such as, for example, the presence of a quantum dot coupled to the wire end \cite{deng2016majorana,moore2018two}, which can be viewed as a strong local inhomogeneity affecting almost all parameters, and position-dependent fluctuations of the semiconductor-superconductor coupling at the interface, which generate not only position-dependent variations of the induced pairing, but, as a result of proximity-induced renormalization \cite{renorm}, produce variations of all relevant parameters (e.g., g-factor, spin-orbit coupling strength, etc.). Indeed, the presence of disorder will generally produce in-gap states that, in the Majorana basis, correspond to pairs of Majorana modes. If (and only if) the system contains the main ingredients for Majorana physics, some of these Majorana pairs will be spatially separated. Assuming that a normal lead is attached to one of the ends of the wire, one will obtain quantized zero bias conductance peaks that are robust against variations of the control parameters (i.e., quantized ZBC islands) whenever (i) the lead couples to a single Majorana mode and (ii) the mode has a certain spatial separation with respect to all other Majorana modes in the system.

In addition to demonstrating that the observation of quantized ZBC islands is selective-enough to exclude false positives generated by trivial low-energy states, which are ubiquitously present in a disordered system, we have shown that systematically generating ZBC maps over large windows in parameter space represents a powerful diagnostic of the disorder present in the hybrid system. If, for example, no quantized island can be observed within the accessible parameter range and after testing multiple nominally-identical structures, there are two possibilities: a) the system is in the extreme disorder limit or b) one of the necessary Majorana ingredients (spin-orbit coupling, Zeeman splitting, superconducting pairing) is absent. Within scenario (a), there is no segment of the wire with effectively-low disorder long-enough to enable the emergence of ps-MMs. We have explicitly tested scenario (b) by  setting the spin-orbit coupling to zero; the resulting ZBC map contains only non-quantized (yellow) islands. 
Upon reducing  disorder from the extreme disorder limit, one should observe (small) quantized islands, which are signatures of (remnant) Majorana physics. The number and size of these islands increase with decreasing disorder and, eventually, they start to coalesce into larger quantized regions. At this point, the disorder may be low-enough to enable additional Majorana  signatures, such as ZBC peaks emerging within the nominally-topological region (and not outside it) and conductance features exhibiting edge-to-edge correlations. 

We emphasize that the zero-bias conductance maps and other similar tools consistent with our proposed approach --  performing large-scale mappings of relevant observable quantities as functions of the control parameters --
can be extremely useful for navigating the unpaved road toward topological Majorana zero modes, as they represent (i) systematic surveys for identifying unique Majorana features (e.g., quantized islands) and (ii) powerful diagnostic tools for evaluating disorder. Specifically, the observation of (even small) quantized Majorana islands should be the next milestone along this road, because it represents the unique Majorana feature that is least susceptible to disorder. Indeed, the emergence of quantized islands only requires ``local'' Majorana physics, which can be realized even in the presence of strong disorder, yet it  demonstrates unambiguously that the hybrid system actually possesses all necessary Majorana ingredients. From this perspective,  looking for edge-to-edge correlations as a signature of MZMs in samples that do not systematically exhibit quantized islands would be premature. 
Finally, we emphasize that the observation of small quantized islands in strongly disordered systems may require sample selection and involves tuning the control parameters 
over extensive regions of the parameter space with high-enough resolution 
to capture the small islands. Nonetheless, since large zero bias conductance peaks of height approaching or exceeding $2e^2/{h}$ have already been observed \cite{nichele2017scaling,zhang2021,Yu_2021}, the obvious next step is demonstrating unambiguous signatures of Majorana physics through the observation of quantized islands, which may still require a reduction of the disorder strength in the hybrid system.      

\begin{acknowledgments}
C. Z. acknowledges support from NSFC (Grant No. 12104043), the fellowship of the China Postdoctoral Science Foundation (Grant No. 2021M690409) and the National Key R$\&$D Program of China (Grant No. 2020YFA0308800). S. T. thanks the NSF 2014157 for support. G. S. acknowledges support from SERB Grant No. SRG/2020/000134. T. D. S. was supported by NSF Grant No. 2014156.
\end{acknowledgments}

\bibliography{biblio.bib, RevPaperREF.bib}
\end{document}